\documentclass[preprint,showpacs,preprintnumbers,amsmath,amssymb]{article}

\usepackage{graphicx}
\usepackage{dcolumn}
\usepackage{bm}

\begin{document}

\title{5-Dimensional Kaluza-Klein Theory with a Source}

\author{Yongge Ma${ }^1$\footnote{yonggema@yahoo.com}\quad and Jun
Wu${ }^{1,2}$\footnote{vividwhy@yahoo.com.cn}\\
\small ${ }^1$ Department of Physics, Beijing Normal University,
Beijing
100875, China \\
\small ${ }^2$ Department of Physics, The Chinese University of
Hong Kong, Hong Kong, China}

\date{\today}

\maketitle

\begin{abstract}
A free test particle in 5-dimensional Kaluza-Klein spacetime will
show its electricity in the reduced 4-dimensional spacetime when
it moves along the fifth dimension. In the light of this
observation, we study the coupling of a 5-dimensional dust field
with the Kaluza-Klein gravity. It turns out that the dust field
can curve the 5-dimensional spacetime in such a way that it
provides exactly the source of the electromagnetic field in the
4-dimensional spacetime after the dimensional reduction.


\end{abstract}

{PACS number(s): 04.50.+h, 04.20.Fy}

\section{Introduction}

   A unified formulation of Einstein's theory of gravitation and
Maxwell's theory of electromagnetism in 4-dimensional spacetime
was first proposed by Kaluza and Klein using 5-dimensional
geometry\cite{Kaluza}\cite{Klein}. The original Kaluza-Klein (K-K)
theory assumed the so-called "\emph{Cylinder Condition}", which
means that there exists a spacelike Killing vector field $\xi^a$
on the 5-dimensional spacetime ($\hat{M}, \hat{g}_{ab}$)
\cite{duff}-\cite{yang}. In addition, Kaluza demands also that
$\xi^a$ is normalized, i.e.,
\begin{equation}
\phi\equiv\hat{g}_{ab}\xi^a\xi^b=1. \label{ansatz}
\end{equation}
Note that the abstract index notation\cite{wald} is employed
throughout the paper and the signature of the 5-metric is of the
convention $(-,+,+,+,+)$. Later research shows that the ansatz
(\ref{ansatz}) may be dropped out and the $\phi$ may play a key
role in the study of cosmology\cite{uzan}-\cite{Mohamm}. Being the
extra dimension, the orbits of $\xi^a$ are geometrically circles.
The physical consideration that any displacement in the usual
physical 4-dimensional spacetime should be orthogonal to the extra
dimension implies that the "physical" 4-dimensional metric should
be defined as:
\begin{equation}
g_{ab}=\hat{g}_{ab}-\phi^{-1}\xi_a\xi_b.\label{gab}
\end{equation}
For practical calculation, it is convenient to take a coordinate
system $\{z^M=(x^\mu, y)|\mu=0,1,2,3\}$ with coordinate basis
$(e_M)^a=\{(e_\mu)^a,(e_5)^a\}$ on $\hat{M}$ adapted to $\xi^a$,
i.e., $(e_5)^a=(\frac{\partial}{\partial y})^a=\xi^a$. Then the
5-metric components $\hat{g}_{MN}$ take the form:

\begin{equation}
\hat{g}_{MN}= \left(%
\begin{array}{cc}
  g_{\mu\nu}+\phi B_\mu B_\nu & \phi B_\mu \\
  \phi B_\nu & \phi \\
\end{array}%
\right), \label{g_{MN}}
\end{equation}
where $\hat{g}_{\mu5}\equiv\phi B_\mu$. So, locally, the
"physical" spacetime can be understood as a 4-manifold $M$ with
the coordinates $\{x^\mu\}$ endowed with the metric $g_{ab}$.

The whole theory is governed by the 5-dimensional Einstein-Hilbert
action:
\begin{equation}
S_G=-\frac 1{2\hat{k}}\int_{\hat{M}}\sqrt{-\hat{g}}\hat{R}.
\label{SG}
\end{equation}
Suppose the range of the fifth coordinate to be $0\leq y\leq L$
and the 4-dimensional gravitational constant to be $k=\hat{k}/L$.
Let $B_\mu=fA_\mu, f^2=2k$, then equation (\ref{SG}) becomes a
coupling action on M as:
\begin{equation}
\hat{S}_G=\int_M\sqrt{-g}\sqrt{\phi}[-\frac 1{2k}R+\frac 14\phi
F_{ab}(A)F^{ab}(A)], \label{Sg4}
\end{equation}
where $R$ is the curvature scalar of $g_{ab}$ on $M$ and
$F_{ab}(A)\equiv 2\partial_{[a}A_{b]}$. Thus, it results in a
4-dimensional gravity $g_{ab}$ coupled to an electromagnetic field
$A_a$ and a scalar field $\phi$. It is clear that, under the
ansatz (\ref{ansatz}), 5-dimensional K-K theory unifies the
Einstein's gravity and the source-free Maxwell's field in the
standard formulism. However, we know in the actual world that any
electromagnetic fields as well as curved geometries are
necessarily caused by some sources. Therefore, to arrive at a
realistic K-K theory, one has to find the source\footnote{After
this paper is published, we notice an earlier work \cite{pav}
which concerns the same issue as ours.} in $\hat{M}$, which is
responsible for the electromagnetic field in $M$.

In section 2, the dynamics of a classical test particle in
5-dimensional K-K spacetime is considered. It is shown that the
motion of the particle along the extra dimension is responsible
for its electricity in the 4-dimensional spacetime. This motives
us to study the coupling of a dust field with the 5-dimensional
gravity in section 3. The corresponding 4-dimensional formulism of
the coupling is obtained. At last, by invoking the ansatz
(\ref{ansatz}), a clear physical explanation of the dust as the
source for the electromagnetic field is provided.

\section{Classical Particle in K-K Spacetime}

The geodesic motion of a classical particle in K-K spacetime has
been studied by many authors \cite{wesson}-\cite{Dahia}. Here we
will sketch the basic idea and try a slightly different
presentation. Let $V^a$ be the 5-velocity of a test particle in
the 5-dimensional spacetime $(\hat{M}, \hat{g}_{ab})$. Then we can
decompose $V^a$ into its projection $u^a$ in the "physical"
spacetime $(M, g_{ab})$ and another vector $u_\bot^a$ which
represents its motion along the fifth dimension, i.e.,
\begin{equation}
u^a\equiv g^a_bV^b=V^\mu(e_\mu)^a-(B_\mu V^\mu)(e_5)^a,
\label{u^a}
\end{equation}
\begin{equation}
u_\perp^a\equiv V^a-u^a=(V^5+B_\mu V^\mu)(e_5)^a, \label{uperp^a}
\end{equation}
where $g^a_b\equiv\delta^a_b-\phi^{-1}\xi^a\xi_b$ is the
projective map which can project any 5-vector into a vector in
$M$. Since $g_{ab}(e_5)^b=0$, the 4-velocity of the particle in
$M$ can be defined as:
\begin{equation}
v^a\equiv \frac{V^\mu(e_\mu)^a}{\sqrt{-V^\mu V_\mu}}, \label{v^a}
\end{equation}
where $V^\mu V_\mu\equiv g_{\mu\nu}V^\mu V^\nu$. The worldline of
a free particle in $(\hat{M}, \hat{g}_{ab})$ is a timelike
geodesic satisfying:
\begin{equation}
\frac{d^2z^M}{d\tau^2}+\hat{\Gamma}^M_{NR}\frac{dz^N}{d\tau}\frac{dz^R}{d\tau}=0,
\label{motion}
\end{equation}
where $\hat{\Gamma}^M_{NR}$ is the Christoffel symbol of
$\hat{g}_{MN}$ and $dz^M/d\tau=V^M$. Eq. (\ref{motion}) can be
rewritten as:
\begin{equation}
\frac{dV^\mu}{d\tau}+\hat{\Gamma}^\mu_{NR}V^NV^R=0,
\label{motion1}
\end{equation}
\begin{equation}
\frac{dV^5}{d\tau}+\hat{\Gamma}^5_{NR}V^NV^R=0. \label{motion2}
\end{equation}
It is not difficult to see that Eq.(\ref{motion2}) implies
\cite{wesson}
\begin{equation}
\frac {dQ}{d\tau}\equiv V^M\partial_M[\phi(B_\rho V^\rho+V^5)]=0.
\label{motion4}
\end{equation}
From Eq.(\ref{g_{MN}}) one knows that the line element of
$\hat{g}_{ab}$ reads
\begin{equation}
d\hat{s}^2=ds^2+\phi(B_\mu dx^\mu+dy)^2, \label{ds}
\end{equation}
where $ds^2$ is the line element of $g_{ab}$. Hence from Eqs.
(\ref{ds}) and (\ref{motion4}) we can obtain the relation of the
particle's proper time $\tau$ in $\hat{M}$ and proper time $t$ in
$M$ as
\begin{equation}
dt=d\tau\sqrt{1+Q^2/\phi}. \label{dt}
\end{equation}
In the light of Eq.(\ref{dt}), one can further show that
Eq.(\ref{motion1}) leads to
\begin{equation}
v^\rho\nabla_\rho v^\mu-\frac {fQ}{\sqrt{1+Q^2/\phi}} F^\mu_\nu(A)
v^\nu-\frac {Q^2}{2(1+Q^2/\phi)\phi^2}(\partial^\mu\phi+v^\mu
v^\nu\partial_\nu\phi)=0, \label{motion5}
\end{equation}
where $\nabla_a$ is the covariant derivative of $g_{ab}$ in $M$.
Since $Q$ is a constant of the motion, we can define a charge:
\begin{equation}
q\equiv\frac{fmQ}{\sqrt{1+Q^2}}, \label{q}
\end{equation}
where $m$ denotes the mass of the classical particle. When we drop
the influence of the scalar field $\phi$ by ansatz (\ref{ansatz}),
Eq.(\ref{motion5}) takes the form:
\begin{equation}
v^b\nabla_b v^a-\frac qmF^a_{ b}(A) v^b=0. \label{motion6}
\end{equation}
It coincides exactly with the equation for the motion of an
electrically charged particle in the electromagnetic field
\cite{liang}, and hence $q$ gets its physical meaning as the
electric charge. Note that the electric charge $q$ defined here is
different from that in Refs. \cite{wesson} and \cite{Milutin}.
Recalling the definition of $u_\bot^a$, we conclude that the
so-called electric charge in four dimensions is just a
manifestation of a particle's motion along the fifth dimension in
K-K spacetime.

\section{A Source of K-K Spacetime}

To look for the source of the electromagnetic field and the curved
K-K spacetime, the result in last section motivates us to consider
a 5-dimensional dust field with 5-velocity $V^a$ in $\hat{M}$. It
is determined by the action:
\begin{equation}
\hat{S}_D=-\int_{\hat{M}}\sqrt{-\hat{g}}\hat{\mu},
 \label{dust5}
\end{equation}
where the proper energy density $\hat{\mu}$ is adjusted to keep
the fluid current vector $j^a\equiv \hat{\mu}V^a$ conserved. The
variation of action (\ref{dust5}) with respect to the flow line of
the fluid yields \cite{Hawking}
\begin{equation}
\hat{\mu}V^b\hat{\nabla}_bV^a=0, \label{dusteq}
\end{equation}
where $\hat{\nabla}_b$ is the covariant derivative of
$\hat{g}_{ab}$ in $\hat{M}$. Hence the flow lines of the dust are
timelike geodesics in the 5-dimensional spacetime. The
energy-momentum tensor of the dust field reads
\begin{equation}
\hat{T}_{ab}=\hat{\mu} V_aV_b. \label{EM}
\end{equation}
Suppose that the dust is reduced to some fluid in $M$. Let $P$ be
a "comoving observer" in $M$, with Eq.(\ref{v^a}) as its
4-velocity. Then $P$ corresponds to an observer $\hat{P}$ in
$\hat{M}$ with 5-velocity
\begin{equation}
\hat{v}^a=\frac{u^a}{\sqrt{-V^\mu V_\mu}},\label{5v}
\end{equation}
where $u^a$ is defined as Eq.(\ref{u^a}). Thus the energy density
of the dust with respect to $\hat{P}$ will be
\begin{equation}
\bar{\mu}=\hat{T}_{ab}\hat{v}^a\hat{v}^b=\hat{\mu}(-V^\mu V_\mu).
\label{barmu}
\end{equation}
Note that we have
\begin{equation}
-1=V^a V_a=V^\mu V_\mu+\phi(V^\rho B_\rho+V^5)^2. \label{uu}
\end{equation}
We now follow the arguments in last section on free particles in
K-K spacetime to endow the electricity to the fluid in $M$.
According to Eq.(\ref{q}), one may define $\bar{\rho}\equiv
f\bar{\mu}Q/\sqrt{1+Q^2}$ to be the "electric charge density" of
the fluid. However, the observer $P$ can only observe 4
dimensions, so the energy density of the fluid, which $P$ actually
measures in $M$, is not $\bar{\mu}$ but $\mu=\bar{\mu}L$.
Correspondingly, the electric charge density in $M$ is
$\rho=L\bar{\rho}$. Similar to Eq.(\ref{motion5}) we can obtain
from Eq.(\ref{dusteq}) the reduced 4-dimensional field equation of
the fluid in $M$ as
\begin{equation}
\mu v^\rho\nabla_\rho v^\mu-\gamma F^\mu_\nu(A) J^\nu-\frac {\mu
Q^2}{2(1+Q^2/\phi)\phi^2}(\partial^\mu\phi+v^\mu
v^\nu\partial_\nu\phi)=0, \label{dusteq2}
\end{equation}
where we have defined $\gamma\equiv \sqrt{\phi(1+Q^2)/(\phi+Q^2)}$
and $J^a=\rho v^a$ is the electric 4-current density.

We now consider the reduction of 5-dimensional Einstein's
equation:
\begin{equation}
\hat{R}_{MN}-\frac 12 \hat{g}_{MN}\hat{R}=\hat{k}\hat{\mu}V_MV_N,
\end{equation}
which is equivalent to
\begin{equation}
\hat{R}_{MN}=\hat{k}\hat{\mu}(V_MV_N+\frac 13 \hat{g}_{MN}).
\label{einstein}
\end{equation}
It is not difficult to show from Eq.(\ref{g_{MN}}) that the
components of the 5-dimensional Ricci tensor $\hat{R}_{ab}$ can be
expressed as \cite{Wehus}:
\begin{equation}
\hat{R}_{55}=\frac 12 k\phi^2F^{\sigma\rho}F_{\sigma\rho}-\frac 12
\nabla^\mu\nabla_\mu\phi+\frac
1{4\phi}(\nabla^\mu\phi)\nabla_\mu\phi, \label{55}
\end{equation}
\begin{equation}
\hat{R}_{\mu 5}=\frac f2(\phi\nabla^\nu F_{\mu\nu}+\frac 32
F_{\mu\nu}\nabla^\nu\phi)+ B_\mu[\frac 12
k\phi^2F^{\sigma\rho}F_{\sigma\rho}-\frac 12
\nabla^\nu\nabla_\nu\phi+\frac
1{4\phi}(\nabla^\nu\phi)\nabla_\nu\phi], \label{mu5}
\end{equation}
\begin{eqnarray}
\hat{R}_{\mu\nu}&=&R_{\mu\nu}-k\phi F^\sigma_\mu
F_{\sigma\nu}-\frac 1{2\phi} \nabla_\mu\nabla_\nu\phi+\frac
1{4\phi^2}(\nabla_\mu\phi)\nabla_\nu\phi \nonumber \\
&&+B_\mu B_\nu[\frac 12 k\phi^2F^{\sigma\rho}F_{\sigma\rho}-\frac
12 \nabla^\sigma\nabla_\sigma\phi+\frac
1{4\phi}(\nabla^\sigma\phi)\nabla_\sigma\phi] \nonumber \\
&&+\frac f2 B_\mu(\phi\nabla^\sigma F_{\nu\sigma}+\frac 32
F_{\nu\sigma}\nabla^\sigma\phi)+\frac f2 B_\nu(\phi\nabla^\sigma
F_{\mu\sigma}+\frac 32 F_{\mu\sigma}\nabla^\sigma\phi).
\label{munu}
\end{eqnarray}
Plunging Eq.(\ref{55}) into Eq.(\ref{einstein}) we obtain a
coupling equation for the matter fields as
\begin{equation}
\frac 12 k\phi^2F^{ab}F_{ab}=\frac 12 \nabla^a\nabla_a\phi-\frac
1{4\phi}(\nabla^a\phi)\nabla_a\phi+k\mu\phi[1-\frac{2\phi}{3(\phi+Q^2)}].
\label{scalar}
\end{equation}
Plunging Eq.(\ref{mu5}) into Eq.(\ref{einstein}) and using
Eq.(\ref{scalar}), we obtain an electromagnetic field equation
with source as
\begin{equation}
\phi\nabla^b F_{ab}+\frac 32F_{ab}\nabla^b\phi=\gamma J_a.
\label{EM}
\end{equation}
Plunging Eq.(\ref{munu}) into Eq.(\ref{einstein}) and using Eqs.
(\ref{scalar}) and (\ref{EM}), we obtain a 4-dimesional Einstein
equation with source as
\begin{equation}
G_{ab}=k[\mu v_a v_b+\phi(F^c_a F_{cb}-\frac 14
g_{ab}F^{cd}F_{cd})+\frac
1{k\sqrt{\phi}}(\nabla_a\nabla_b\sqrt{\phi}-g_{ab}\nabla^c\nabla_c\sqrt{\phi})],
\label{gravity}
\end{equation}
where $G_{ab}$ is the Einstein tensor of $g_{ab}$.

Now we turn to the variational principle. In the light of
Eq.(\ref{uu}), action (\ref{dust5}) can be written as:
\begin{equation}
\begin{array}{lll}
\hat{S}_D&=&\int_{\hat{M}}\sqrt{-\hat{g}}\frac{\bar{\mu}}{V^\mu
V_\mu}\\
&=&-\int_{\hat{M}}\sqrt{-\hat{g}}\bar{\mu}(1+\frac{\phi(V^\rho
B_\rho+V^5)^2}{V^\mu V_\mu})\\
&=&-\int_{\hat{M}}\sqrt{-\hat{g}}\bar{\mu}(1-\frac{QB_\lambda
v^\lambda}{\sqrt{-V^\mu V_\mu}}+\frac{QV^5}{V^\mu V_\mu}),
\end{array}
\label{dust51}
\end{equation}
which is reduced to a 4-dimensional action for the fluid on $M$
as:
\begin{equation}
\hat{S}_D=-\int_M\sqrt{-g}\sqrt{\phi}(\mu-\gamma J^\mu
A_\mu-\frac{\gamma\rho V^5}{f\sqrt{1+Q^2/\phi}}). \label{dust}
\end{equation}
Thus the total action of K-K gravity coupled with a dust field
reads:
\begin{equation}
\hat{S}_G+\hat{S}_D=\int_M\sqrt{-g}\sqrt{\phi}[-\frac 1{2k}R+\frac
14\phi F_{\mu\nu}(A)F^{\mu\nu}(A)-\mu+\gamma J^\mu
A_\mu+\frac{\gamma\rho V^5}{f\sqrt{1+Q^2/\phi}}]. \label{S}
\end{equation}
Recall that the K-K spacetime admits a killing vector field
$\xi^a$, which will correspond to a conservative vector:
\begin{equation}
P^a=\hat{T}^{ab}\xi_b=\hat{\mu}V^aV^b(e_5)_b =\frac
\gamma{Lf}[J^\mu(e_\mu)^a+\frac{\rho
V^5(e_5)^a}{\sqrt{1+Q^2/\phi}}]. \label{P}
\end{equation}
It follows that $\sqrt{-\hat{g}}P^a$ and hence the components
$\sqrt{-\hat{g}}\gamma\rho V^5/(f\sqrt{1+Q^2/\phi})$ and
$\sqrt{-\hat{g}}\gamma J^\mu$ are invariant under the variation
with respect to the metric components $\hat{g}_{MN}$ (or
$g_{\mu\nu}, \phi$ and $A_\mu$)\cite{Hawking}. Therefore the last
term in action (\ref{S}) can be neglected if one only concerns the
fields equations equivalent to the 5-dimensional Einstein
equations. We thus get an action on $M$ as
\begin{equation}
S=\int_M\sqrt{-g}\sqrt{\phi}[-\frac 1{2k}R+\frac 14\phi
F_{ab}(A)F^{ab}(A)-\mu+\gamma J^a A_a], \label{S2}
\end{equation}
which coincides with the 4-dimensional coupled action of gravity,
charged fluid, electromagnetic field and a scalar field. Its
variations with respect to $\phi$, $A^\mu$, and $g^{\mu\nu}$ yield
respectively Eqs. (\ref{scalar}), (\ref{EM}) and (\ref{gravity})
exactly. Note that the energy density $\mu$ contains also the
variational information of $\hat{\mu}$ with respect to $\phi$
whence the last two terms in action (\ref{S}) are invariant.

\section{Discussion}

To reveal the physical meaning of the coupled dust field, we
invoke Kaluza's ansatz (\ref{ansatz}). Then the field equations
(\ref{dusteq2}), (\ref{scalar}), (\ref{EM}) and (\ref{gravity})
become respectively
\begin{equation}
\mu v^b\nabla_b v^a=F^a_{ b} J^b, \label{dusteq3}
\end{equation}
\begin{equation}
\frac 12 F^{ab}F_{ab}=k\mu[1-\frac 2{3(1+Q^2)}], \label{scalar2}
\end{equation}
\begin{equation}
\nabla^b F_{ab}=J_a, \label{EM2}
\end{equation}
\begin{equation}
G_{ab}=k(T^{(D)}_{ab}+T^{(E)}_{ab}), \label{gravity2}
\end{equation}
where $T^{(D)}_{ab}\equiv\mu v_a v_b$ and $T^{(E)}_{ab}\equiv
F^c_a F_{cb}-\frac 14 g_{ab}F^{cd}F_{cd}$ are respectively the
usual energy-momentum tensors of the dust and the electromagnetic
fields. Subject to Eq.(\ref{scalar2}), Eqs. (\ref{dusteq3}),
(\ref{EM2}) and (\ref{gravity2}) are respectively the standard
4-dimensional Lorentz, Maxwell and Einstein equations for the
coupling of charged fluid, electromagnetic field and gravity.
Therefore, the 4-dimensional manifestation of a dust field in K-K
spacetime can be understood as the source of the reduced
electromagnetic field in the "physical" spacetime, provided the
dust has motion along the fifth dimension. Thus we have found the
source for the electromagnetic field as well as the curved
5-dimensional spacetime. Note that a test particle can only show
the effects of a spacetime on itself by its motion, but the
particle does not affect the background spacetime. Hence the
well-known fact about a free particle in a given K-K spacetime
could not ensure our result, which concerns the effects of a dust
field on the 5-dimensional spacetime. This is certainly a
non-trivial physical result, although the motivation is obvious.
It further completes the physical explanation of the electric
charge as the motion of a free particle along the extra dimension
in K-K theory.

The physical meaning of the dust field is obvious also from the
viewpoint of the action. If we had imposed ansatz (\ref{ansatz})
into action (\ref{S2}) directly, it would become
\begin{equation}
S=\int_M\sqrt{-g}[-\frac 1{2k}R+\frac 14
F_{ab}(A)F^{ab}(A)-\mu+J^a A_a], \label{S3}
\end{equation}
which coincides with the standard 4-dimensional coupled action of
gravity, electromagnetic field and charged fluid. This action
would result in the field equations (\ref{gravity2}) and
(\ref{EM2}) without the restrictive equation (\ref{scalar2}).

\section*{ Acknowledgments}

This work is supported in part by NSFC (grant 10205002) and YSRF
for ROCS, SEM.


\begin{thebibliography}{99}

\bibitem{Kaluza} T. Kaluza, Sitzungsber. Preuss. Akad. Wiss. Phys. Mat. Klasse, 966 (1921).

\bibitem{Klein} O. Klein, Z. Phys. {\bf 37}, 895 (1926).

\bibitem{duff} M. J. Duff and B.E.W. Nilsson, Phys. Rep. {\bf 130}, 1 (1986).

\bibitem{Overduin} J.M. Overduin and P.S. Wesson, Phys. Rep. {\bf 283}, 305(1997).

\bibitem{yang} X. Yang, Y. Ma, J. Shao, and W. Zhou, Phys. Rev. D{\bf
68}, 024006 (2003).

\bibitem{wald}  R.M. Wald, General relativity, (The University of Chicago Press,
1984).

\bibitem{uzan} J-P. Uzan, Rev. Mod. Phys. {\bf 75}, 403 (2003).

\bibitem{BR} P.J.E. Beebles and B. Ratra, Rev. Mod. Phys. {\bf 75}, 599 (2003).

\bibitem{Wehus} I.K. Wehus and F. Ravndal, "Dynamics of scalar field in 5-dimensional Kaluza-Klein
theory". Hep-ph/0210292.

\bibitem{Mohamm} N. Mohammedi, Phys. Rev. D{\bf 65}, 104018
(2002).

\bibitem{pav} N. Mankoc-Borstnik and M. Pavsic, IL Nuovo Cimento
{\bf 99}, 489 (1988).

\bibitem{wesson} P.S. Wesson, Space-Time-Matter: Modern
Kaluza-Klein Theory, (World Scientific Publishing, 1999).

\bibitem{Milutin} M. Blagojevic, Gravitation and Gauge Symmetry, (IOP Publishing, 2002).

\bibitem{Seahra} S.S. Seahra, Phys. Rev. D{\bf 65}, 124004 (2002).

\bibitem{Leon} J. Ponce de Leon, Grav. Cosmol. {\bf 8}, 272 (2002).

\bibitem{Dahia} F. Dahia, E.M. Monte, and C. Romero, Mod. Phys. Lett. A{\bf 18}, 1773
(2003).

\bibitem{liang} C. Liang, Introductory Differential Geometry and General
Relativity I, II (Beijing Normal University Press, 2000, in
Chinese).

\bibitem{Hawking} S.W. Hawking and G.F.R. Ellis, The Large Scale Structure of
Space-Time, (Cambridge University Press, 1980).

\end{thebibliography}
\end{document}